\documentclass[runningheads]{llncs}
\usepackage{times}
\usepackage{epsfig}
\usepackage{graphicx}
\usepackage{amsmath}
\usepackage{amssymb}
\usepackage[utf8]{inputenc} % allow utf-8 input
\usepackage[T1]{fontenc}    % use 8-bit T1 fonts
\usepackage[hidelinks]{hyperref}
\hypersetup{
  colorlinks   = true, %Colours links instead of ugly boxes
  urlcolor     = blue, %Colour for external hyperlinks
  linkcolor    = blue, %Colour of internal links
  citecolor   = red %Colour of citations
}
\usepackage{url}            % simple URL typesetting
\usepackage{booktabs}       % professional-quality tables
\usepackage{amsfonts}       % blackboard math symbols
\usepackage{amsmath}
\usepackage{nicefrac}       % compact symbols for 1/2, etc.
\usepackage{microtype}      % microtypography
\usepackage{tabularx}
\usepackage{multirow}
\usepackage{appendix}
\usepackage{lipsum}
\usepackage{caption}
\usepackage{subcaption}% Used for displaying a sample figure. If possible, figure files should

\begin{document}
\title{Hierarchical 3D Feature Learning for Pancreas Segmentation}
\author{Federica Proietto Salanitri \inst{1} \and Giovanni Bellitto \inst{1} \and Ismail Irmakci \inst{2, 3}
\and Simone Palazzo \inst{1} \and Ulas Bagci \inst{3} \and Concetto Spampinato \inst{1}}
%\institute{Anonymous for double-blind review}
%\author{First Author\inst{1}\orcidID{0000-1111-2222-3333} \and
%Second Author\inst{2,3}\orcidID{1111-2222-3333-4444} \and
%Third Author\inst{3}\orcidID{2222--3333-4444-5555}}
\institute{
PeRCeiVe Lab, University of Catania, Italy. \and 
CE, Ege University, Izmir, Turkey \and
Department of Radiology and BME, Northwestern University, Chicago, IL, USA
}
%Springer Heidelberg, Tiergartenstr. 17, 69121 Heidelberg, Germany

%
\maketitle              % typeset the header of the contribution
\begin{abstract}
We propose a novel 3D fully convolutional deep network for automated pancreas segmentation from both MRI and CT scans. More specifically, the proposed model consists of a 3D encoder that learns to extract volume features at different scales; features taken at different points of the encoder hierarchy are then sent to multiple 3D decoders that individually predict intermediate segmentation maps. Finally, all segmentation maps are combined to obtain a unique detailed segmentation mask. 
We test our model on both CT and MRI imaging data: the publicly available NIH Pancreas-CT dataset (consisting of 82 contrast-enhanced CTs)  and a private MRI dataset (consisting of 40 MRI scans). Experimental results show that our model outperforms existing methods on CT pancreas segmentation, obtaining an average Dice score of about 88\%, and yields promising segmentation performance on a very challenging MRI data set (average Dice score is about 77\%). Additional control experiments demonstrate that the achieved performance is due to the combination of our 3D fully-convolutional deep network and the hierarchical representation decoding, thus substantiating our architectural design.

\end{abstract}
\section{Introduction}
\label{sec:intro}
Pancreatic cancer is  a growing public health concern worldwide. In 2021, an estimated 60,430 new cases of pancreatic cancer will be diagnosed in the US and 48,220 people will die from this disease~\cite{society2021cancer}. Early detection of pancreas cancer~\cite{oberstein2013pancreatic} is very hard and options in treatment are very limited. Radiology imaging and automated image analysis play key roles in diagnosis, prognosis, treatment, and intervention of pancreatic diseases; thus, there is a strong, unmet, need for computer aided analysis tools supporting these tasks. The first step in such analysis is to automate the medical image segmentation procedures, since manual segmentation (current standard) is tedious, prone to error, and it is not practical in routine clinical evaluation of the diseases~\cite {european2019radiologist}. Beyond the known challenges of medical image segmentation problems, pancreas is one of the most difficult organs to segment despite the recent advances in deep segmentation models. 

 Computed tomography (CT) and magnetic resonance imaging (MRI) are the two most common modalities for pancreas imaging. CT is the modality of choice for pancreatic cancer at the moment, while MRI is mostly used for finding other pancreatic diseases including cysts and diabetes. Compared to CT, MRI has advantages such as the lack of ionizing radiation, better resolution and soft tissue contrast. However, MRI has other unique difficulties, including field inhomogeneity, non-standard intensity distributions due to variations in scanners, patients, field strengths, and high similarity in pancreas and non-pancreas tissue densities.\\ Image-based pancreas analysis is by itself a challenging task. Shapes and sizes greatly vary across different patients, making it difficult to use robust priors for improving the delineation procedures. Intensity similarities to non-pancreatic tissues, and smooth or invisible boundaries (due to resolution limitations of medical scanners) are other challenges that need to be addressed in a successful segmentation method. Moreover, in presence of a cyst, tumor, or other abnormalities in pancreases, segmentation algorithms may easily fail to delineate correct boundaries.

To address these challenges, in this work we propose a novel 3D fully convolutional encoder-decoder network with hierarchical multi-scale feature learning, for general, fully-automated pancreas segmentation applicable to CT and MRI scans. Major contributions of this study are the following:
\begin{itemize}
  \item Our segmentation network is unique in the sense that it is volumetric, learns to extract 3D volume features at different scales, and decodes features hierarchically, leading to improved segmentation results;
  \item We show the efficacy of our work both on CT and MRI scans. Our architecture successfully extracts pancreases from CT and MRI with high accuracy, obtaining new state-of-the-art results on a publicly-available CT benchmark and first-ever volumetric pancreas segmentation from MRI in the literature.
  \item Our work on MRI pancreas segmentation is an important application contribution, due to the very limited published research on this task using MRI data with deep learning. It is our belief that our method provides a significant state-of-the-art baseline to be compared with for further MRI pancreas research. 
\end{itemize}

\section{Related Work}
\label{sec:related}
Following the success of deep learning methods applied in medical image segmentation, researchers have recently shown an increasing interest in pancreas segmentation, in order to support physicians in early stage diagnosis for pancreas cancer. Although this application field is still in its infancy --- also due to variabilities in texture, size and imaging contrast --- a line of promising approaches has been proposed in the literature, mainly on CT scans~\cite{cai2017improving,khosravan2019pan,8880606,8937496,8692647,roth2015deeporgan,roth2016spatial,roth2018spatial,9098473,wang2021pancreas,8578962,zhou2017fixed}. We here describe the most significant ones which relate to our proposed model.\\ In~\cite{roth2016spatial}, a two-stage cascaded approach for pancreas localization and pancreas segmentation is proposed. In the first stage, the method localizes the pancreas in the entire 3D CT scan, providing a reliable bounding box for a more refined segmentation step, based on an efficient application of holistically-nested convolutional networks (HNNs) on the three views of pancreas CT image. Per-pixel probability maps are then fused to produce a 3D bounding box of the pancreas.
Projective adversarial networks~\cite{khosravan2019pan} incorporate high-level 3D information through 2D projections and introduce an attention module that supports a selective integration of global information from the segmentation module to an adversarial network.
More recently, ~\cite{wang2021pancreas} proposes a dual-input v-mesh fully-convolutional network, which receives original CT scans and images processed by contrast-specific graph-based visual saliency, in order to enhance the soft tissue contrast and highlight differences among local regions in abdominal CT scans. 

All of the above works tackle the problem of pancreas segmentation on CT scans. However, as already mentioned, MRI acquisitions have several advantages over CT --- most importantly, fewer risks to the patients. On the other hand, MRI pancreas segmentation presents additional challenges to automated visual analysis. For this reason and others (e.g., the lack of public benchmarks), very few works have addressed pancreas segmentation on MRI data: to the best of our knowledge, the major attempts are~\cite{asaturyan2019morphological,cai2017improving,cai2016pancreas}. In~\cite{cai2016pancreas}, two CNN models are combined to perform, respectively, tissue detection and boundary detection; the results are provided as input to a conditional random field (CRF) for final segmentation.  
In~\cite{asaturyan2019morphological}, an algorithmic approach based on hand-crafted features is proposed, employing an ad-hoc multi-stage pipeline: contrast enhancement within coarsely detected pancreas regions is applied to differentiate between pancreatic and surrounding tissue; 3D segmentation and edge detection through max-flow and min-cuts approach and structured forest are performed; finally, non-pancreatic contours are removed via morphological operations on area, structure and connectivity.

\section{Method}
\label{sec:method}
\begin{figure*}
    \centering
    \begin{subfigure}{0.25\textwidth}
    \centering
    \includegraphics[height=2cm]{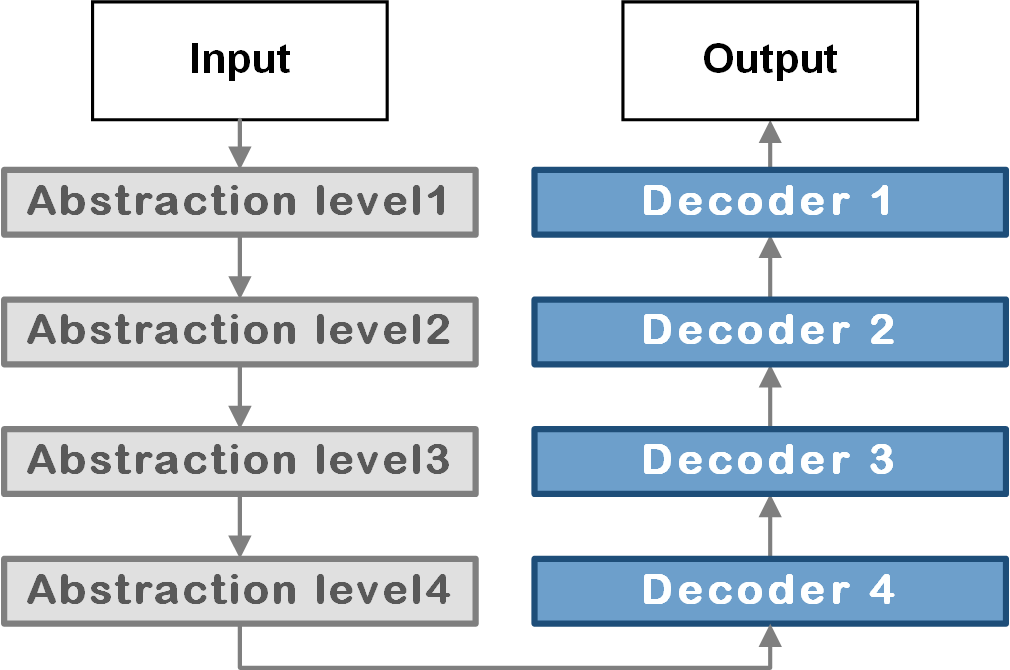}
    \caption{encoder-decoder}
    \label{fig:1a}
    \end{subfigure}
    \hspace{0.4cm}
    \begin{subfigure}{0.3\textwidth}
    \centering
    \includegraphics[height=2cm]{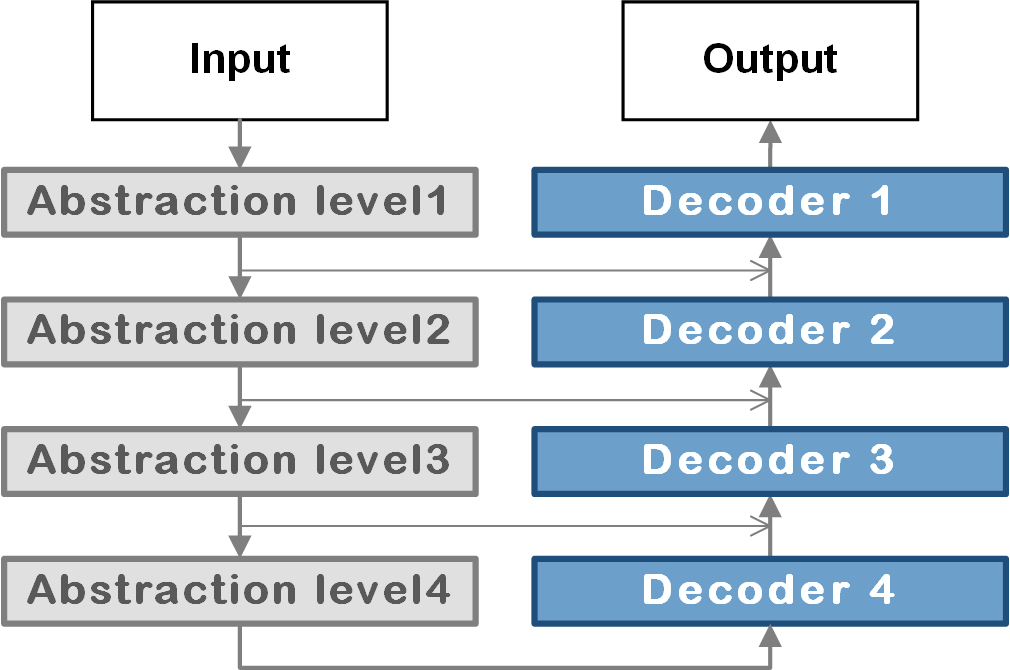}
    \caption{skip-connection}
    \label{fig:1b}
    \end{subfigure}
    \begin{subfigure}{0.35\textwidth}
    \centering
    \includegraphics[height=2cm]{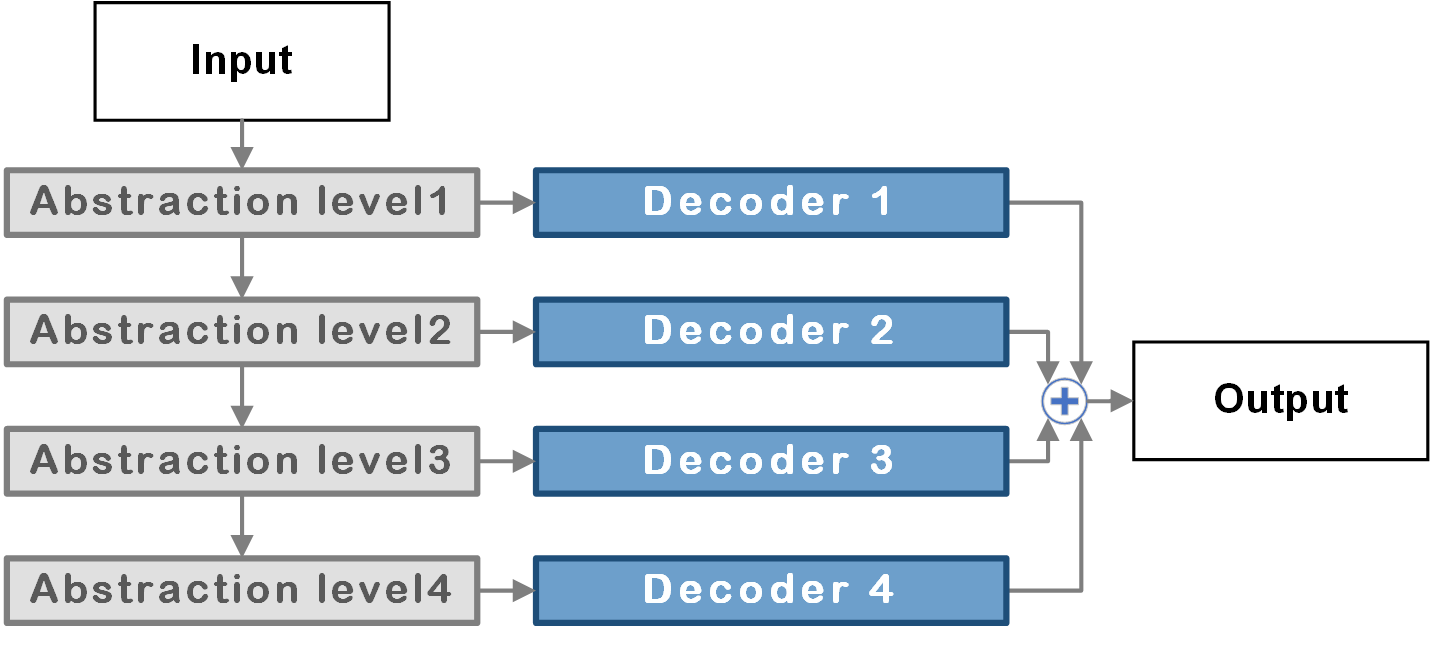}
    \caption{hierarchical decoding}
    \label{fig:1c}
    \end{subfigure}
    \caption{A comparison between our proposed architecture and other types of networks used for segmentation: (a) standard encoder--decoder architecture; (b) encoder--decoder architecture with  skip connections; (c) encoder--hierarchical decoder architecture (ours).}
    \label{fig:hierarchical}
\end{figure*}

\begin{figure*}[h!]
    \centering
    \includegraphics[width = 0.70\textwidth]{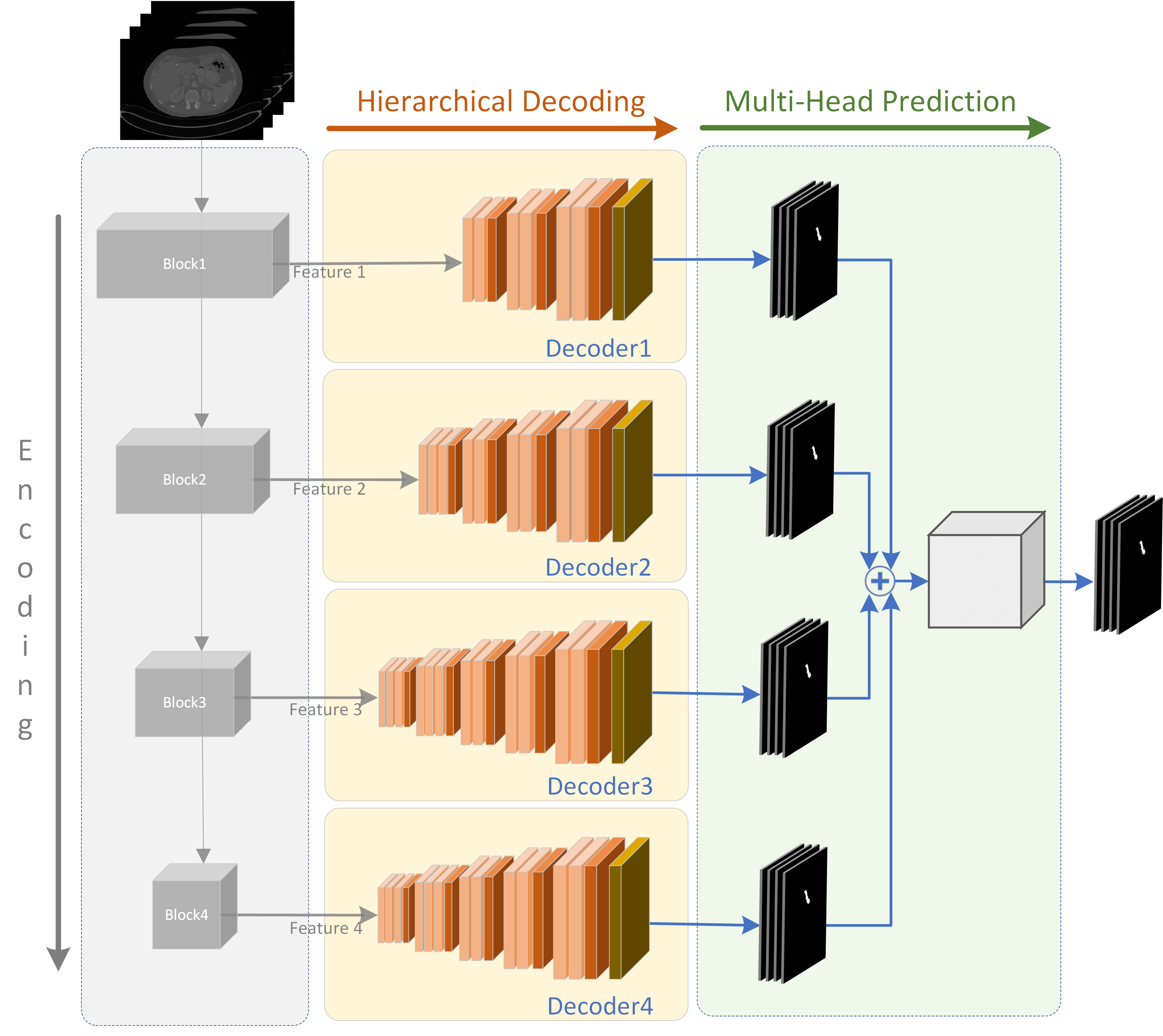}
    \caption{\textit{PanKNet} architecture: the encoding path extracts aggregated volumetric features, while the decoding path predicts four different intermediate segmentation masks (coarse to fine). Finally, intermediate segmentations are integrated into a detailed output mask. }
    \label{fig:PanKNet}
\end{figure*}

Our 3D fully-convolutional pancreas segmentation model --- \textit{PankNet} --- is based on an encoder-decoder architecture; however, unlike standard encoder-decoder schemes with a single decoding path (see Fig.~\ref{fig:1a}), we have parallel decoders at different abstraction levels, generating multiple intermediate segmentation maps (Fig.~\ref{fig:1c}). Hierarchical decoding is also fundamentally different from using skip connections (Fig.~\ref{fig:1b}), since these have the purpose to ease gradient flow and forward low-level features for output reconstruction, while our multiple decoders aim to extract local and global dependencies. 
The detailed architecture is shown in Fig.~\ref{fig:PanKNet}: the input data (either CT or MRI volume) is first processed by the encoder stream of the model which aggregates volumetric features at different abstraction levels. These features are then given as input to different decoder streams, each generating a segmentation mask volume. 
All intermediate masks are concatenated along the channel dimension and finally merged through a convolutional layer in order to predict the final segmentation mask for all input slices. 

\subsection{Volume feature encoding}
\label{sub:encoding}
The model's encoder performs aggregation of volumetric features from the input data. It is based on S3D~\cite{xie2018rethinking}, a network originally proposed for action recognition using 3D spatial and temporal separable 3D convolution layers, pretrained on the Kinetics Dataset~\cite{kay2017kinetics}. We use the pretrained network, similarly to other works~\cite{cai2016pancreas,khosravan2019pan,lalonde2019inn}, to ease convergence given the limited training data we have from both CT and MRI datasets.
Our encoder processes $D=48$ slices from an input scan by progressively aggregating volumetric cues down to a more compact representation of size 1024$\times\frac{W}{8} \times\frac{H}{32}\times\frac{D}{32}$ (channels $\times$ width $\times$ height $\times$ depth). Features at the bottleneck and at the outputs of the second, third and fourth pooling layers are fed to separate decoders, described in the following section, to implement our hierarchical decoding strategy.

The proposed approach can be easily adapted to different encoder architectures. Thus, we additionally design a lightweight variant of our \textit{PanKNet} network by replacing the S3D-based encoder with an encoder based on MobileNetV2~\cite{sandler2018mobilenetv2}, where 2D convolutions are replaced with 3D ones through inflation. In particular, the 2D kernels are replicated along the third dimension, and the values of the weights are divided by the number of replications as proposed in~\cite{carreira2017quo}. In this case, as input to the decoders, we select the output of the second, third, fourth and sixth \textit{bottleneck} blocks of MobileNetV2, providing a more compact feature map of size 160$\times\frac{W}{16}\times\frac{H}{32}\times\frac{D}{32}$.
This lightweight variant has 10 times fewer parameters (2.5 millions of parameters, 9.33 MB) and  than the S3D counterpart (25.6 millions of parameters, 97.88 MB). 

\subsection{Hierarchical Decoding}
\label{sub:decoding}
Our hierarchical decoding strategy employs features at different points of the encoder stream to generate intermediate segmentation masks that aim to capture and combine fine segmentation (derived from decoders of deeper features) to coarse segmentation (derived from decoders of initial features).  We include four decoders: each one processes a set of volumetric features taken from the corresponding level in the encoder stack and performs segmentation on the input volume (see Fig.~\ref{fig:PanKNet}, yellow blocks). 
Each decoder consists of a cascade of upsampling blocks, depending on the size of the input feature map: decoders operating on deeper features require less blocks to recover the original input size. Each upsampling block contains a 3D convolutional block (convolutional layer + batch normalization + ReLU), one or two 3D separable convolutional blocks, and a trilinear upsample layer. As last layer, a pointwise 3D convolution outputs a volume with size 2 $\times W\times H\times D$, where $W$, $H$ and $D$ are the same as the input volume.  

\subsection{Pancreas Segmentation}
\label{sub:sal_pred}
Intermediate segmentation maps predicted by each of the model's decoders are combined into a global mask. In particular, the four intermediate maps are concatenated into a 8$\times W \times H \times D$ tensor, which then goes through a last layer performing a voxel-wise convolution to generate a single segmentation map of size $2 \times W \times H \times D$.

The whole model (encoder, hierarchical decoders and output layer) is trained end-to-end using a hierarchical Dice loss~\cite{7785132} between ground-truth mask, intermediate generated masks and the output segmentation mask.
Formally, given the predicted output segmentation masks $\mathbf{S}_v$ for the input volume, the four maps $\mathbf{\hat{S}}_{v^i}$ %with $i=1,...,4$
estimated by the decoders, and the ground-truth segmentation maps $\mathbf{G}_v$ for the input data, the \textit{segmentation loss} $\mathcal{L}_s$ is:

\begin{equation}\label{loss}
 \mathcal{L}_s\left( \mathbf{S}_v, \mathbf{\hat{S}}_{v^i}, \mathbf{G}_v \right) = \sum_{i=1}^4 \frac{2 \sum_{j} {\hat{S}}_{v^{i,j}}  {G}_{v^j}}{\sum_j {\hat{S}}_{v^{i,j}}^2 + \sum_j {G}_{v^{j}}^2} + \frac{2 \sum_{j} {S}_{v^{j}}  {G}_{v^j}}{\sum_j {S}_{v^{j}}^2 + \sum_j {G}_{v^{j}}^2} 
\end{equation}
where index $i$ iterates over the four intermediate maps and index $j$ iterates over voxels.

\section{Experiments}
\label{sec:exp}

\subsection{Dataset}
\label{sec:dataset}
We evaluate the accuracy of our proposed deep segmentation method in both CT and MRI modalities. For the former, we use the publicly available NIH Pancreas-CT dataset, which is the most used pancreas segmentation dataset for benchmarking~\cite{roth2015deeporgan}. This dataset includes 82 abdominal contrast-enhanced 3D CT scans. The resolution of the CT scans is 512 $\times$ 512 $\times$ $Z$, with $Z$ (between 181 and 466) indicating the number of slices along the transverse axis. Voxel spacing ranges from 0.5 mm to 1 mm. More details on this dataset are available in~\cite{roth2015deeporgan}.

In our experiments with MRI data, we use 40 in-house collected T2-weighted MRI scans from 40 patients, who have either IPMN (intraductal papillary mucinous neoplasm) cysts detected in their pancreases or invasive pancreatic ductal carcinoma. Two expert radiologists annotated pancreases manually and consensus segmentation masks were generated at the end of the ground-truth labeling procedure with agreement. MRI images were resized (in the transverse plane) to 256 x 256 pixels, with voxel spacing of varying from 0.468 mm to 1.406 mm. To minimize uncertainties in MRI scans, we applied a set of pre-processing steps: N4 bias field correction followed by an edge-preserving Gaussian smoothing, and intensity standardization procedure to standardize MRI scans across patients, scanners, and time. 

\subsection{Training and evaluation procedure}
\label{sec:training}

We apply the same training procedure for the two datasets, with the only difference regarding how model backbones are pre-trained. On the NIH Pancreas-CT dataset, we pre-train S3D on Kinetics~\cite{kay2017kinetics} and MobileNetV2 on ImageNet~\cite{deng2009imagenet} with weight inflation; on our MRI data, \textit{Pancreas-MRI}, we employ the backbones pre-trained on the CT task.

Input CT and MRI scans are re-oriented using the RAS axes convention for consistency. We then perform voxel resampling through trilinear interpolation in order to have isotropic (1 mm) voxel spacing, and normalize the values of each scan between 0 and 1. During training, data augmentation is performed with random horizontal flipping, random 90-degrees rotation and random crops of size 128$\times$128$\times$48 (in RAS coordinates). We minimize our multi-part Dice loss with mini-batch gradient descent using the Adam optimizer (learning rate: 0.001) and batch size 8, for a total of 3000 epochs.

At inference time, we compute output segmentation masks by running a sliding window routine over an entire input scan, using 256$\times$256$\times$48 windows overlapping by 25\%. Voxel labels from overlapping segmentations are obtained by averaging the set of predictions. For evaluation, we carry out 4-fold cross-validation. At each iteration, the set of training folds is further split into the actual training set and a validation set, that is used to select the epoch at which Dice score on the test fold is reported.
As metrics for quantitative evaluation, we employ: \textit{Dice score coefficient} (DSC), \textit{Positive Predictive Value} (PPV) and \textit{Sensitivity}.

Experiments are performed on an NVIDIA Quadro P6000 GPU. The proposed approach was implemented in PyTorch and MONAI; all code will be publicly released.  

\begin{table}[h!]
\centering
\begin{tabular}{ccccccccc}
\toprule
\textbf{\textbf{Method}} & & \multicolumn{3}{c}{\textbf{DSC}}& & \textbf{PPV} & &\textbf{SENS}\\
\cmidrule{1-1} \cmidrule{3-5} \cmidrule{7-7} \cmidrule{9-9}
& & Avg & Max & Min & & & & \\
\cmidrule{3-5}
Roth et al.~\cite{roth2015deeporgan} && 71.42 $\pm$ 10.11 \hspace{1em}& 86.29 \hspace{1em}& 23.99 && -- && -- \\
Roth et al.~\cite{roth2016spatial}   && 78.01 $\pm$ 8.20 \hspace{1em}& 88.65 \hspace{1em}& 34.11 && -- && -- \\
Roth et al.~\cite{roth2018spatial}   && 81.27 $\pm$ 6.27 \hspace{1em}& 88.96 \hspace{1em}& 50.69 && -- && -- \\
Zhou et al.~\cite{zhou2017fixed}     && 82.37 $\pm$ 5.68 \hspace{1em}& 90.85 \hspace{1em}& 62.43 && -- && -- \\
Cai et al.~\cite{cai2017improving}   && 82.40 $\pm$ 6.70 \hspace{1em}& 90.10 \hspace{1em}& 60.00 && -- && -- \\
Li et al. (2019)~\cite{8880606}      && 83.50 $\pm$ 6.20 \hspace{1em}& -- \hspace{1em}& -- && 84.50 $\pm$ 6.90  && 83.70 $\pm$ 10.40\\
Liu et al. (2020)~\cite{8937496}     && 84.10 $\pm$ 4.90 \hspace{1em}& -- \hspace{1em}& -- && 83.60 $\pm$ 5.90 && 85.30 $\pm$ 08.20\\
You et al.~\cite{8578962}            && 84.50 $\pm$ 4.97 \hspace{1em}& 91.02 \hspace{1em}& 62.81 && -- && -- \\
Khosravan et al.~\cite{khosravan2019pan} && 85.53 $\pm$ 1.23 \hspace{1em}& 88.71 \hspace{1em}& 83.20 && -- && -- \\
Wang et al. (2020)~\cite{9098473}    && 85.90 $\pm$ 3.40 \hspace{1em}& -- \hspace{1em}& -- && -- && --\\
Man et al. (2019)~\cite{8692647}     && 86.90 $\pm$ 4.90 \hspace{1em}& -- \hspace{1em}& -- && -- && --\\
Wang et al.~\cite{wang2021pancreas}  && 87.04 $\pm$ 6.80 \hspace{1em}& -- \hspace{1em}& -- && \textbf{89.50} $\pm$ \textbf{5.80} && 87.70 $\pm$ 7.90\\
\midrule
PanKNet\textsubscript{Light} && 87.13 $\pm$ 4.58 & 93.49 & \textit{72.77} && 86.85 $\pm$ 6.52 && 88.48 $\pm$ 5.12\\
PanKNet && \textbf{88.01} $\pm$ \textbf{4.74} & \textbf{93.84} & 70.62 && \textit{88.25} $\pm$ \textit{5.45} && \textbf{88.69} $\pm$ \textbf{5.99}\\
\bottomrule
\end{tabular}
\caption{Comparison of \textit{PanKNet} against multiple state-of-the-art models for pancreas segmentation on NIH Pancreas-CT dataset using 4-fold cross-validation. \textbf{Best performance in bold}, \textit{second best in italic}.}
\label{tab:ct_segmentation}
\end{table}

\subsection{Results}
\label{sec:results}
We first test our model (as well as its lightweight variant) on the NIH Pancreas-CT dataset and compare it to existing methods (which share our evaluation strategy with 4-fold cross-validation), namely, \cite{cai2017improving,khosravan2019pan,8880606,8937496,8692647,roth2015deeporgan,roth2016spatial,roth2018spatial,9098473,wang2021pancreas,8578962,zhou2017fixed}. 
Summarized in Table~\ref{tab:ct_segmentation}, our results indicate that \textit{PanKNet} outperforms existing methods over different metrics. Note that PanKNet does not require any auxiliary regularization networks~\cite{khosravan2019pan}, nor additional inputs~\cite{wang2021pancreas}, nor upstream pancreas localization module~\cite{8692647}. 
Remarkably, even the lightweight variant of PanKNet yields accuracy comparable to the full model, while outperforming existing models, showing that the choice of the backbone is not as important as the overall employed hierarchical architecture. 
The best trade-off between accuracy and computational resources for CT pancreas segmentation is represented by PanKNet\textsubscript{Light}, whose memory occupation is about 10 MB compared to about 100 MB of PanKNet, but with very similar performance. \\
We then test our model on pancreas segmentation from MRI data. In this case, we compare the 3D-UNet, proposed in~\cite{kerfoot2018left}, pre-trained on the NIH Pancreas-CT dataset and fine-tuned on our MRI dataset. Furthermore, we add to this evaluation some control experiments to show the effectiveness of the designed architecture. Consequently, we define as baseline our encoder-decoder architecture without hierarchical decoding strategy, decoding only the features at the model's bottleneck. 
Results in Table~\ref{tab:mri_segmentation} indicate that both PanKNet variants outperform the state-of-the-art 3D U-Net model~\cite{kerfoot2018left}. The baseline (with either backbones) also performs better than 3D U-Net model~\cite{kerfoot2018left} demonstrating that even our 3D fully convolutional network, ablated from the hierarchical decoding, is effective for MRI pancreas segmentation. Adding hierarchical decoding leads to enhanced segmentation performance, especially on DSC and PPV. Different from  CT segmentation and from baseline models, PanKNet largely outperforms its lightweight counterpart, demonstrating that MRI pancreas segmentation is far more complex and challenging than CT segmentation and calls for high-capacity networks to be solved. 

\begin{figure}
    \centering
    \rotatebox{90}{CT Segmentation}
    \includegraphics[width=0.43\textwidth]{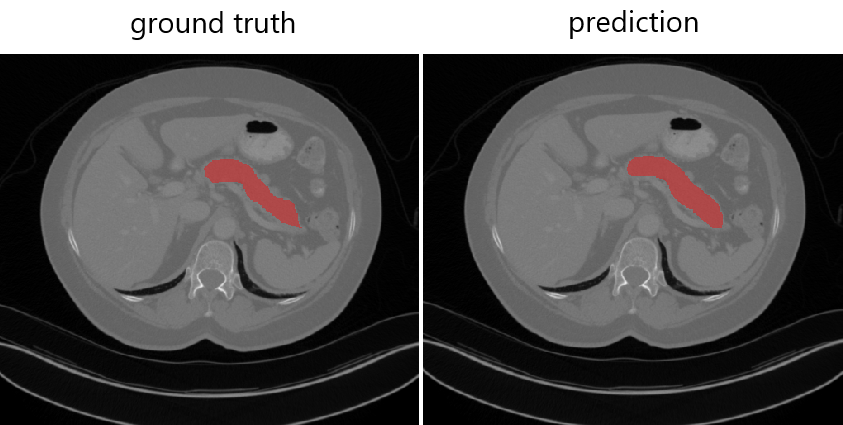}
    \includegraphics[width=0.43\textwidth]{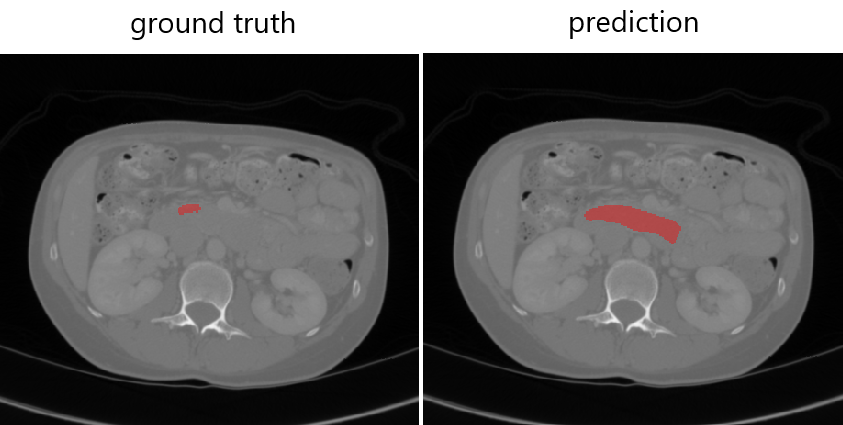}\\
    \rotatebox{90}{MRI Segmentation}
    \includegraphics[width=0.43\textwidth]{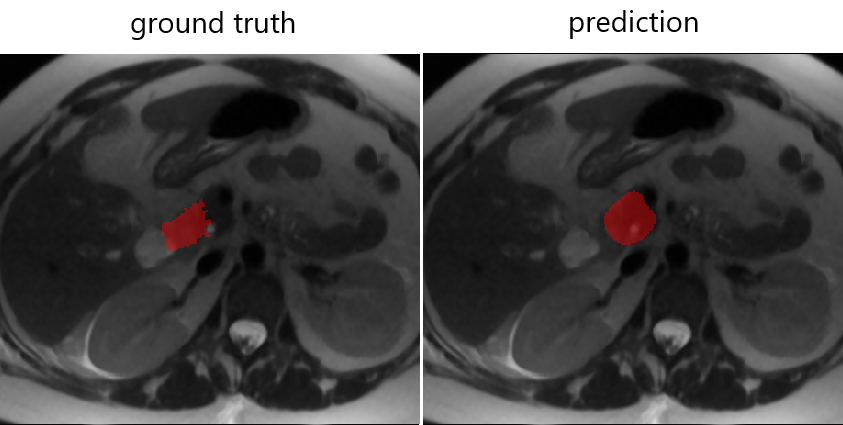}
    \includegraphics[width=0.43\textwidth]{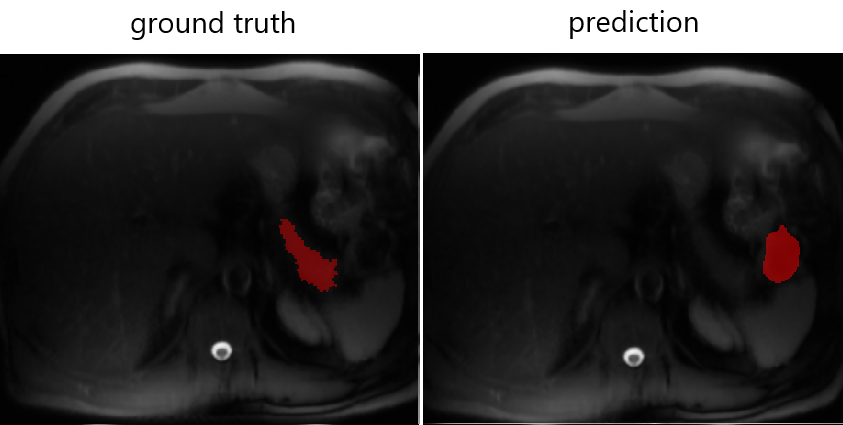}\\
    \begin{minipage}[c]{1\textwidth}
    \vspace{0.1cm}
    \hspace{1.3cm} Segmentation at the highest DSC \hspace{1.1cm} Segmentation at the lowest DSC
    \end{minipage}
    \caption{Segmentation masks at the highest  (left column) and lowest Dice score (right column) on NIH Pancreas-CT (first row) and Pancreas-MRI dataset (second row).} \vspace{-0.5cm}
    \label{fig:segmentation_masks}
\end{figure}

Example segmentation masks, corresponding to the highest and lowest Dice scores reported in Tables~\ref{tab:ct_segmentation} and \ref{tab:mri_segmentation} for CT and MRI pancreas segmentation, are illustrated in Fig.~\ref{fig:segmentation_masks}.

\begin{table}[h!]
\centering
\begin{tabular}{ccccccccc}
\toprule
Method& & \multicolumn{3}{c}{\textbf{DSC}}& & \textbf{PPV} & &\textbf{SENS}\\
\cmidrule{1-1} \cmidrule{3-5} \cmidrule{7-7} \cmidrule{9-9}
& & Avg & Max & Min & & & & \\
\cmidrule{3-5}
3D-UNet~\cite{kerfoot2018left} && 65.05 $\pm$ 9.17 \hspace{1em}& 84.58 \hspace{1em}& 49.80 && 61.55 $\pm$ 7.55 && 74.42 $\pm$ 13.99 \\
Baseline\textsubscript{Light} && 69.17 $\pm$ 8.10 \hspace{1em}& 83.86 \hspace{1em}& 49.92 && 64.64 $\pm$ 7.49 && 84.19 $\pm$ 11.72 \\
Baseline && 65.16 $\pm$ 9.11 \hspace{1em}& 84.00 \hspace{1em}& 49.49 && 61.92 $\pm$ 8.22 && 75.22 $\pm$ 12.46 \\
PanKNet\textsubscript{Light} && 72.96 $\pm$ 10.33 \hspace{1em}& 88.54 \hspace{1em}& 49.90 && 71.39 $\pm$ 11.21 && 79.76 $\pm$ 11.53 \\
PanKNet   && \textbf{77.46 $\pm$ 08.62} \hspace{1em}& \textbf{89.07} \hspace{1em}& \textbf{52.30} && \textbf{76.63 $\pm$  08.66} && \textbf{80.91 $\pm$ 10.51} \\
\bottomrule
\end{tabular}
\caption{Segmentation performance on Pancreas-MRI dataset (4-fold CV).}
\label{tab:mri_segmentation}
\end{table}

\section{Conclusion}
\label{sec:conclusion}
In this study, we propose a novel 3D fully-convolutional network for pancreas segmentation from MRI and CT scans. Our proposed deep network aims at learning and combining multi-scale features, namely a hierarchical decoding strategy, to generate intermediate segmentation masks for a coarse-to-fine segmentation process. The intermediate masks, capturing fine details, are derived from decoders of deeper features while coarse segmentation details are derived from decoders of initial features. We evaluated the efficacy of our method (a) on CT scans from the publicly available NIH CT-Pancreas benchmark, and obtained a new state of the art Dice score \textbf{88.01\%}, outperforming all previous methods; and (b) on MRI scans, obtaining a Dice score of \textbf{77.46\%}, which can be used as a baseline for future works on MRI pancreas segmentation. Noting that MRI pancreas segmentation methods are extremely limited due to the challenging nature of the problem, our study offers a fresh insight into MRI analysis of pancreas from a fully automated volumetric segmentation strategy. PanKNet is tested for pancreas segmentation, but its architecture is general and can be applied to any 3D object segmentation problem in medical domain.

%\bibliographystyle{splncs04}
%\bibliography{biblio}

\begin{thebibliography}{10}
\providecommand{\url}[1]{\texttt{#1}}
\providecommand{\urlprefix}{URL }
\providecommand{\doi}[1]{https://doi.org/#1}

\bibitem{asaturyan2019morphological}
Asaturyan, H., Gligorievski, A., Villarini, B.: Morphological and multi-level
  geometrical descriptor analysis in ct and mri volumes for automatic pancreas
  segmentation. Computerized Medical Imaging and Graphics  \textbf{75},  1--13
  (2019)

\bibitem{cai2017improving}
Cai, J., Lu, L., Xie, Y., Xing, F., Yang, L.: Improving deep pancreas
  segmentation in ct and mri images via recurrent neural contextual learning
  and direct loss function. arXiv preprint arXiv:1707.04912  (2017)

\bibitem{cai2016pancreas}
Cai, J., Lu, L., Zhang, Z., Xing, F., Yang, L., Yin, Q.: Pancreas segmentation
  in mri using graph-based decision fusion on convolutional neural networks.
  In: International Conference on Medical Image Computing and Computer-Assisted
  Intervention. pp. 442--450. Springer (2016)

\bibitem{carreira2017quo}
Carreira, J., Zisserman, A.: Quo vadis, action recognition? a new model and the
  kinetics dataset. In: CVPR. pp. 6299--6308 (2017)

\bibitem{deng2009imagenet}
Deng, J., Dong, W., Socher, R., Li, L., Li, K., Li, F.: Imagenet: {A}
  large-scale hierarchical image database. In: Computer Society Conference on
  Computer Vision and Pattern Recognition. pp. 248--255 (2009).
  \doi{10.1109/CVPR.2009.5206848},
  \url{https://doi.org/10.1109/CVPR.2009.5206848}

\bibitem{kay2017kinetics}
Kay, W., Carreira, J., Simonyan, K., Zhang, B., Hillier, C., Vijayanarasimhan,
  S., Viola, F., Green, T., Back, T., Natsev, P., et~al.: The kinetics human
  action video dataset. arXiv preprint arXiv:1705.06950  (2017)

\bibitem{kerfoot2018left}
Kerfoot, E., Clough, J., Oksuz, I., Lee, J., King, A.P., Schnabel, J.A.:
  Left-ventricle quantification using residual u-net. In: International
  Workshop on Statistical Atlases and Computational Models of the Heart. pp.
  371--380. Springer (2018)

\bibitem{khosravan2019pan}
Khosravan, N., Mortazi, A., Wallace, M., Bagci, U.: Pan: Projective adversarial
  network for medical image segmentation. In: International Conference on
  Medical Image Computing and Computer-Assisted Intervention. pp. 68--76.
  Springer (2019)

\bibitem{lalonde2019inn}
LaLonde, R., Tanner, I., Nikiforaki, K., Papadakis, G.Z., Kandel, P., Bolan,
  C.W., Wallace, M.B., Bagci, U.: Inn: inflated neural networks for ipmn
  diagnosis. In: International Conference on Medical Image Computing and
  Computer-Assisted Intervention. pp. 101--109. Springer (2019)

\bibitem{8880606}
{Li}, H., {Lü}, Q., {Chen}, G., {Huang}, T., {Dong}, Z.: Convergence of
  distributed accelerated algorithm over unbalanced directed networks. IEEE
  Transactions on Systems, Man, and Cybernetics: Systems pp. 1--12 (2019).
  \doi{10.1109/TSMC.2019.2946287}

\bibitem{8937496}
{Liu}, S., {Yuan}, X., {Hu}, R., {Liang}, S., {Feng}, S., {Ai}, Y., {Zhang},
  Y.: Automatic pancreas segmentation via coarse location and ensemble
  learning. IEEE Access  \textbf{8},  2906--2914 (2020).
  \doi{10.1109/ACCESS.2019.2961125}

\bibitem{8692647}
{Man}, Y., {Huang}, Y., {Feng}, J., {Li}, X., {Wu}, F.: Deep q learning driven
  ct pancreas segmentation with geometry-aware u-net. IEEE Transactions on
  Medical Imaging  \textbf{38}(8),  1971--1980 (2019).
  \doi{10.1109/TMI.2019.2911588}

\bibitem{7785132}
{Milletari}, F., {Navab}, N., {Ahmadi}, S.: V-net: Fully convolutional neural
  networks for volumetric medical image segmentation. In: 2016 Fourth
  International Conference on 3D Vision (3DV). pp. 565--571 (2016).
  \doi{10.1109/3DV.2016.79}

\bibitem{oberstein2013pancreatic}
Oberstein, P.E., Olive, K.P.: Pancreatic cancer: why is it so hard to treat?
  Therapeutic advances in gastroenterology  \textbf{6}(4),  321--337 (2013)

\bibitem{roth2015deeporgan}
Roth, H.R., Lu, L., Farag, A., Shin, H.C., Liu, J., Turkbey, E.B., Summers,
  R.M.: Deeporgan: Multi-level deep convolutional networks for automated
  pancreas segmentation. In: International conference on medical image
  computing and computer-assisted intervention. pp. 556--564. Springer (2015)

\bibitem{roth2016spatial}
Roth, H.R., Lu, L., Farag, A., Sohn, A., Summers, R.M.: Spatial aggregation of
  holistically-nested networks for automated pancreas segmentation. In:
  International conference on medical image computing and computer-assisted
  intervention. pp. 451--459. Springer (2016)

\bibitem{roth2018spatial}
Roth, H.R., Lu, L., Lay, N., Harrison, A.P., Farag, A., Sohn, A., Summers,
  R.M.: Spatial aggregation of holistically-nested convolutional neural
  networks for automated pancreas localization and segmentation. Medical image
  analysis  \textbf{45},  94--107 (2018)

\bibitem{sandler2018mobilenetv2}
Sandler, M., Howard, A., Zhu, M., Zhmoginov, A., Chen, L.C.: Mobilenetv2:
  Inverted residuals and linear bottlenecks. In: IEEE Conference on Computer
  Vision and Pattern Recognition. pp. 4510--4520 (2018)

\bibitem{society2021cancer}
Society, A.C.: Cancer facts \& figures. American Cancer Society  (2021)

\bibitem{european2019radiologist}
of~Radiology (ESR) communications@ myesr. org Emanuele Neri Nandita~de Souza
  Adrian Brady Angel Alberich Bayarri Christoph D. Becker Francesca Coppola
  Jacob~Visser, E.S.: What the radiologist should know about artificial
  intelligence--an esr white paper. Insights into imaging  \textbf{10}, ~1--8
  (2019)

\bibitem{9098473}
{Wang}, W., {Song}, Q., {Feng}, R., {Chen}, T., {Chen}, J., {Chen}, D.Z., {Wu},
  J.: A fully 3d cascaded framework for pancreas segmentation. In: 2020 IEEE
  17th International Symposium on Biomedical Imaging (ISBI). pp. 207--211
  (2020). \doi{10.1109/ISBI45749.2020.9098473}

\bibitem{wang2021pancreas}
Wang, Y., Gong, G., Kong, D., Li, Q., Dai, J., Zhang, H., Qu, J., Liu, X., Xue,
  J.: Pancreas segmentation using a dual-input v-mesh network. Medical Image
  Analysis  \textbf{69},  101958 (2021)

\bibitem{xie2018rethinking}
Xie, S., Sun, C., Huang, J., Tu, Z., Murphy, K.: Rethinking spatiotemporal
  feature learning: Speed-accuracy trade-offs in video classification. In:
  ECCV. pp. 305--321 (2018)

\bibitem{8578962}
{Yu}, Q., {Xie}, L., {Wang}, Y., {Zhou}, Y., {Fishman}, E.K., {Yuille}, A.L.:
  Recurrent saliency transformation network: Incorporating multi-stage visual
  cues for small organ segmentation. In: 2018 IEEE/CVF Conference on Computer
  Vision and Pattern Recognition. pp. 8280--8289 (2018).
  \doi{10.1109/CVPR.2018.00864}

\bibitem{zhou2017fixed}
Zhou, Y., Xie, L., Shen, W., Wang, Y., Fishman, E.K., Yuille, A.L.: A
  fixed-point model for pancreas segmentation in abdominal ct scans. In:
  International conference on medical image computing and computer-assisted
  intervention. pp. 693--701. Springer (2017)

\end{thebibliography}
%\printbibliography

\end{document}